\documentclass{article}
\usepackage{authblk}
\usepackage{amsthm}
\usepackage{amssymb}
\usepackage{amsmath}
\usepackage{amsfonts}
\usepackage{wrapfig}
\usepackage{pgfpages}
\theoremstyle{plain}
\newtheorem{theorem}{Theorem}
\newtheorem{conclusion}{Conclusion}

\title{Users' traffic on two-sided Internet platforms. Qualitative dynamics}
\author{Victoria Rayskin \thanks{The revision of some models and their analysis as well as the manuscript preparation were supported by ARO grant \# W911NF-19-1-0399.} \\victoria.rayskin@tufts.edu
}

\begin{document}
\maketitle
\begin{abstract}
Internet platforms' traffic defines important characteristics of platforms, such as price of services, advertisements, speed of operations. 
One can estimate the traffic with the traditional time series models like ARIMA, Holt-Winters, functional and kernel regressions. When using these methods, we usually want to smooth-out noise and remove various external effects in the data and obtain short-term predictions of processes. However, these models do not necessarily help us to understand the underlying mechanism and the tendencies of the processes.

In this article, we discuss the dynamical system approach to the modeling, which is designed to discover  the underlying mechanism and the qualitative properties of the system's  phase portrait. We show how to reconstruct the governing differential equations from data. The external effects are modeled as system's parameters (initial conditions). Utilizing this new approach, we construct the models for the volume of users, interacting through Internet platforms, such as Amazon.com, Homes.mil or Wikipedia.org. Then, we perform qualitative analysis of the system's phase portrait and discuss the main characteristics of the platforms. 
\end{abstract}
\section{Introduction}\label{section-intro}

Internet platforms become one of the most popular objects for our daily activities.
There are many examples of interactions through platforms: health exchange platforms connect health insurance providers \& subscribers, bidding platforms connect auction buyers \& sellers, real estate platforms connect renters \& home owners, the `1000 genomes' platform connects multi-disciplinary researchers. The first three of these are examples of two-sided platforms and the last example is the example of multi-sided platform (\cite{RT}, \cite{EPA}). 

Users' traffic is one of the most important characteristics of platforms. For example, it helps the platform owners to price the platform's services, to predict platform's crashes due to high traffic and to plan platform development. 

We discuss the dynamical systems' approach for the study of the volume of users interacting through platforms. 

Our goal is to understand the platform-specific rules that govern the Internet platform dynamics.
The new dynamical systems approach developed for platform analysis in \cite{R1} and \cite{R2} makes it now possible to develop strategies for increasing platforms' efficiency. 

Furthermore,  in the study of multi-sided platforms (which becomes increasingly important), dynamical systems are very well suited for a high number of platform sides.

Finally, prediction of a platform's future behavior may be sensitive to noise, external effects and platform's policies. The dynamical systems approach allows one to associate changes in the external world with the changes in the trajectories' initial conditions. Also, one can study regions of stability and other fundamental  qualitative concepts less sensitive to external effects.

In Section~\ref{section-statistics}  we discuss a method for deriving differential equations of the process, from the data. There are numerous techniques for the short-term predictions of the state variables of the process. We demonstrate (Section~\ref{section-vf-vs-flow}) that approximation of the vector field instead of the state variables may result in more accurate predictions and in better understanding of the underlying process. 

The reconstructed vector field allows to perform qualitative analysis of the process and to understand the tendency of the process (see Sections~\ref{section-approximation-examples} and \ref{section-Internet}). Moreover, we can switch between different trajectories of the dynamical system to reflect the changes in the external conditions.

In  Section~\ref{section-approximation-examples} we provide examples of the reconstructed dynamical systems and perform qualitative analysis of the phase portraits.

In Section~\ref{section-Internet} we discuss specific Internet platform models and the characteristics of the mechanism that governs the dynamics of these models. In particular, we compare the properties of the ``seller-buyer'' type of platforms and platforms like Wikipedia. We note that the competition between ``sellers'' as well as competition between ``buyers'' restricts grows of the number of users of a "seller-buyer" type of platform; while Wikipedia's popularity can grow (if there are no "wars" between contributors).

\section{Vector field estimation vs. flow estimation}\label{section-vf-vs-flow}
Many laws of nature and human activities can be modeled with the help of differential equations. Frequently, as illustrated with the next two examples, it is easier to observe and discover  these equations than their solutions.

At the end of 16th century, Galileo Galilei had studied the Law of Free Fall. According to his pupil, Vincenzo Viviany, Galileo had dropped balls from the Learning Tower of Pisa and recorded the data of his experiment.  

Galileo Galilei found that the velocity of the falling objects is proportional to their time of descent. He formulated the Law of Free Fall as the linear differential equation: $$x'(t)=-gt.$$ 
The trajectory (solution) of this equation is quadratic: $$x(t)=h_0-\frac{g}{2}t^2.$$
Even though the tale about the Learning Tower of Pisa may be apocryphal, it would be harder to discover from the data, that the hight $x$ is proportional to the square of the elapsed time $t$, then to discover the linear equation for the velocity.

If we turn our attention to the study of the physical systems, meteorological dynamics or chemical reactions that can be described with the help of the Lorentz equations, we will face complexity of the chaotic behavior. The differential equations of this process are rather simple,  quadratic, and easily discoverable with the modern techniques:
$$
\left\{
\begin{array}{l}
x_1'=10(x_2-x_1),\\
x_2'=x_1(28-x_3)-x_2,\\
x_3'=x_1x_3-\frac{8}{3} x_3
\end{array}
\right.
$$
 However, if we try to describe the Lorentz attractor trajectory, corresponding to the solution of these equations, we would not be able to write an explicit equation and would not be able to achieve high accuracy with approximations.
\begin{wrapfigure}{r}{.5\textwidth}
\centering
\includegraphics[scale=0.22]{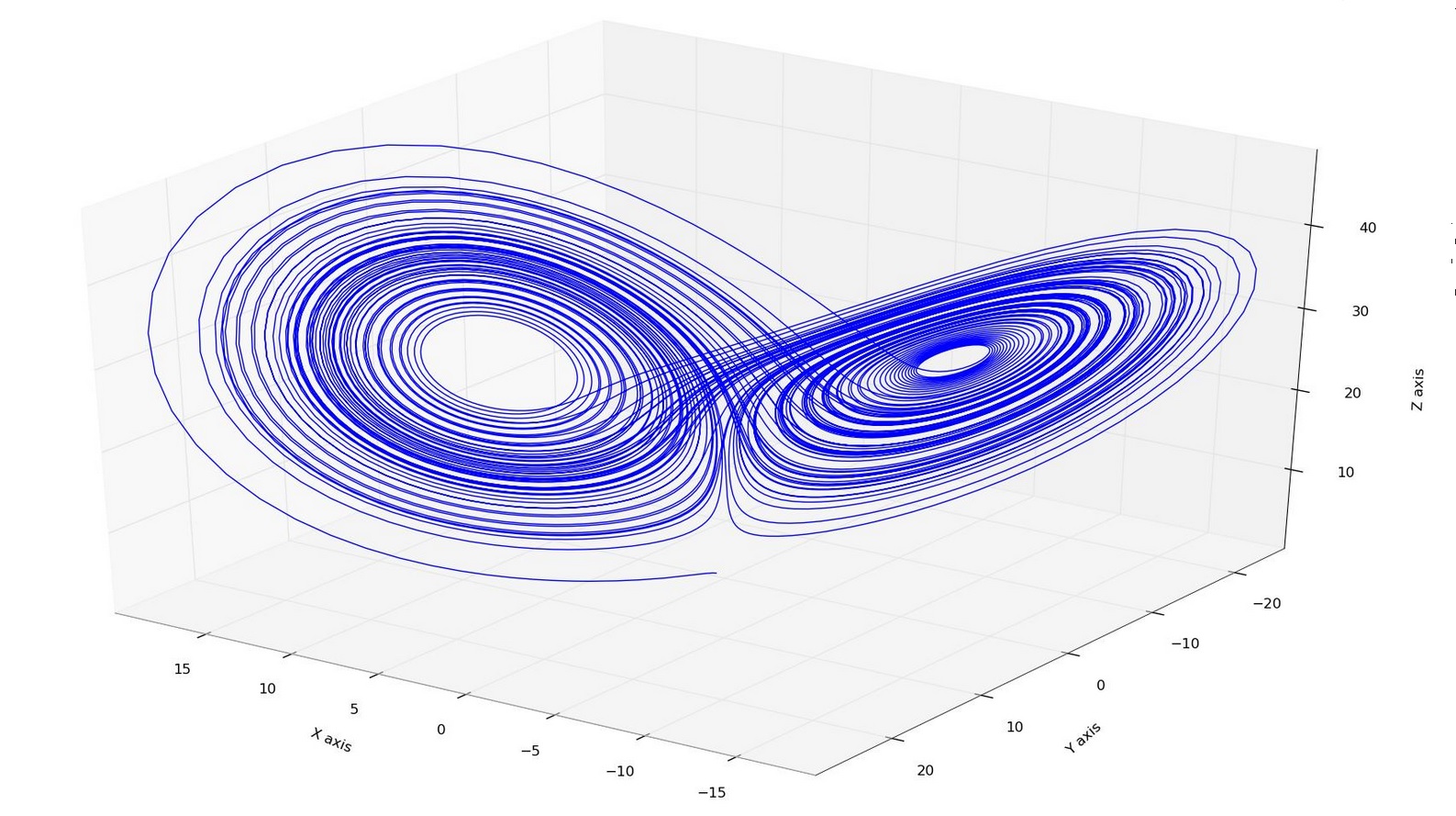}
\caption{Lorentz attractor}
\end{wrapfigure}
Many processes, for which the state variables' equations cannot be explicitly written due to their complexity, can be approximated with simple differential equations, which have very few non-linear terms.

Our goal is to study the complex processes with the help of differential equations that can be fitted into the data, generated by the processes.

This approach allows to understand the general laws governing the process, to perform qualitative analysis of the system's phase portrait and, depending on the external state, to choose one of the initial value's trajectory for the short term prediction of the process. 

Comparing the short term prediction discussed above with the traditional Time Series model's prediction, we found that the former method gives us higher accuracy.

Consider, for instance, the quarterly data for 1955-1985 Gross National Product (GNP) and money supply (M1).  
\begin{figure}
\centering
 \includegraphics[scale=.4]{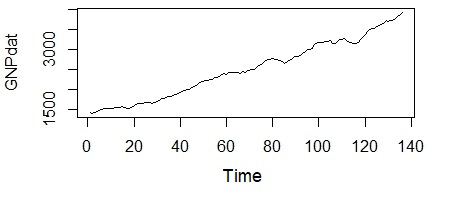} 
\includegraphics[scale=.4]{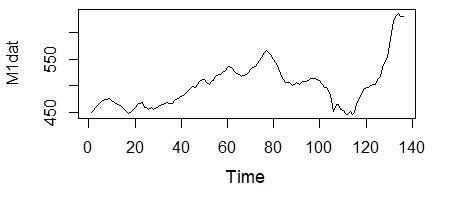} 
\caption{GNP and Money Supply quarterly data for 1955-1985}
\end{figure}
First, let us use a traditional Time Series Model. Since M1 and GNP influence each other, we can employ Vector Auto-Regression. P.Cowpertwait and A.Metcalfe (\cite{CM}) suggest to use VAR of order 3 (VAR(3)).

Let us now assume that there exists "GNP-M1-law", and try to discover it from the data. This "law" is affected by various external factors, for example, government regulations. These externalities will shift the process from one trajectory to another in the phase portrait, which satisfy different initial conditions of the "GNP-M1-law", but the main governing forces of the law will remain unchanged. This is similar to the body, moving in accordance with the Law of Free Fall, but slightly shifted from one initial value's solution to another by a strong wind. 

Using the quarterly data, we can approximate the "GNP-M1-law" with the following dynamical system:
\begin{equation}\label{eqn-gnp-m1}
\left\{
\begin{array}{l}
\begin{aligned}
GNP' = & -855.98 - .0114\cdot GNP + 3.79\cdot M1 \\
          & + .000011 \cdot GNP^2 - .0038 \cdot M1^2 + .00012 \cdot GNP \cdot M1
\\
M1'\  = & -164.69 - .0239 \cdot GNP + .689 \cdot M1 \\
        & + .000008 \cdot GNP^2 - .0005 \cdot M1^2 - .00003 \cdot GNP \cdot M1
\end{aligned}
\end{array}
\right.
\end{equation}

The vector field defined by \eqref{eqn-gnp-m1} can be used for approximation of the solution. For the short term prediction, Root Mean Square Error of this approximation is smaller than the Root Mean Square Error of the VAR(3) method of \cite{CM}.

With the dynamical system approach, we also obtain the equations, which help us to understand how M1 and GNP influence each other.

\section{Techniques of reconstruction of the vector field from the data}\label{section-statistics}
How to reconstruct dynamical systems from data? We will follow the ideas of the SINDy method, developed by Brunton, Proctor, Kutz (\cite{BPK}). However, instead of trying to reconstruct one single trajectory, we will assume that the data comes from many different trajectories (because the process is influenced by external effects and jumps from one initial value solution to another one). We will try to reconstruct the vector field for some region of these trajectories.

Accordingly, we will use LASSO, Ridge Regression, or a similar sparse regression for the vector field approximation, using the data points $\bar{x}(t_j)$. Here $t_j=1,...,p$ are the moments of times, when the state variables $\bar{x}(t_j)$ were recorded.

We will approximate the vector field $\bar{x'}(t_j):=\left(\bar{x}(t_j)-\bar{x}(t_{j-1})\right)/\left( t_j-t_{j-1}\right)$ by a linear combination of functions from the library 
\begin{equation*}\label{eqn-library}
{\bf L(X})=\{L_k(\bar{x}(t_j)) \}_{k,j=1}^{K,p}.
\end{equation*} 
(Here, $K$ is the number of functions included in the library ${\bf L}$.)

More specifically, for an $N$-dimensional system, we want to find the optimal values $A=\{\hat{a}_{ki}\}_{k,i=1}^{K,N}$ s.th. ${\bf L(X}) \cdot \widehat{A}$ approximates $({\bf X'})^T$:
\begin{equation*}\label{eqn-stat-approximation}
x'_i(t_j) \approx \widehat{x'_i(t_j)}=\sum_k L_k(\bar{x}(t_j))\cdot \hat{a}_{ki}.
\end{equation*}

The library ${\bf L}$ of "candidate" functions  may consist of homogenious monoms, trigonometric and/or other simple functions. The matrix $\widehat{A}$ is sparse, because we need only a few terms for the vector field approximation. 

\section{Examples of the vector field reconstruction}\label{section-approximation-examples}
In this section, we consider several examples of differential equations reconstructed (precisely or approximately) from data, using the method, discussed in Section~\ref{section-statistics}.  These examples are related to the dynamics of the volume of users on Internet platforms, which we will present in Section~\ref{section-Internet}.

Our first example will be based on the data, simulated by the system of differential equations
\begin{equation}\label{sqrt-equations-system}
\left\{
\begin{array}{l}
x' = -x +\sqrt{y},\\
y' = -y +\sqrt{x}.
\end{array}
\right. 
\end{equation}

\begin{figure}
\centering
\includegraphics[scale=0.4]{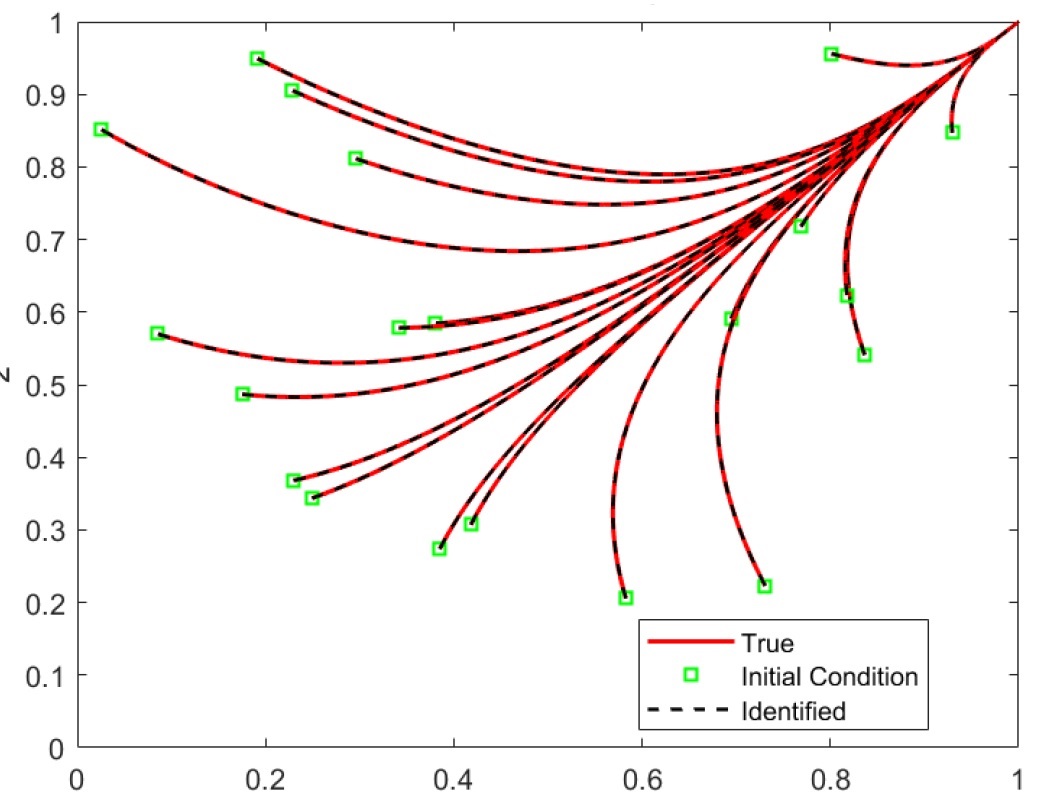}
\caption{True and identified trajectories, defined by the Equations~\eqref{sqrt-equations-system}. All trajectories in the basin of attraction $[0,1]^2\setminus {0}$ converge to $(1,1)$. The trajectories' initial conditions are shown as green squares.}
\label{pic-flow-sqrt}
\end{figure}

For the library ${\bf L}$, we will use monoms of order $\leq 5$ and $\sqrt{x}, \sqrt{y}$. The LASSO regression is capable of reconstructing the Equations~\eqref{sqrt-equations-system} precisely. The reconstructed dynamical system
can be analyzed, the global properties can be described, and the short term predictions can be performed starting at any initial condition. The phase portrait of this system, which has 1 basin of attraction with the attracting fixed point $(1,1)$ is shown in Figure~\ref{pic-flow-sqrt}.

\begin{figure}
\centering
\includegraphics[scale=0.4]{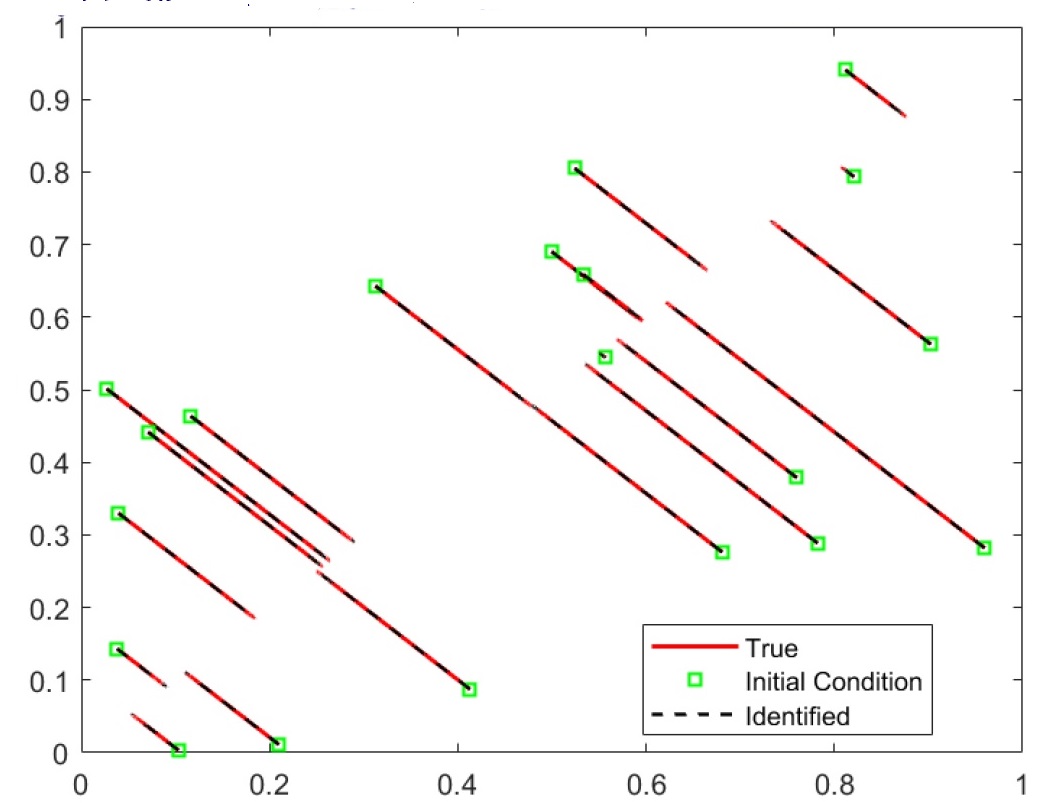}
\caption{True and identified trajectories, defined by the Equations~\eqref{linear-equations-system}. All trajectories converge to the diagonal $y=x$. The trajectories' initial conditions are shown as green squares.}
\label{pic-flow-linear}
\end{figure}

Next, we will simulate the data with the help of the linear system
\begin{equation}\label{linear-equations-system}
\left\{
\begin{array}{l}
x' = -x +y,\\
y' = -y +x.
\end{array}
\right. 
\end{equation}
Theoretically, this system is very simple. However, it has the whole diagonal of stationary points. This singularity may cause problems with the traditional techniques for predicting future values of the process. However, if we try to reconstruct the differential equations from the simulated data, using the same library ${\bf L}$ as in the last example, we can recover the equations precisely, and obtain the simple picture of the linear flow (Figure~\ref{pic-flow-linear}).

Clearly, once we have discovered the linear nature of the process, it is very easy to analyze this flow and to predict the future of the process.

In practice, the functions which we include in the library $L$ do not coincide with the functions (or other mechanisms) that generate the real process. In the next example, we will simulate the data with the help of the system, which consists of the functions not included in our library:
\begin{equation}\label{smoothed-step-system}
\left\{
\begin{array}{l}
x' = -x +y - \frac{1}{4\pi}\sin(4\pi y),\\
y' = -y +x -\frac{1}{4\pi}\sin(4\pi x)
\end{array}
\right. 
\end{equation}
We will use the library of monoms of order $\leq 5$, $\sin(nx)$ and $\cos(nx)$, $n=1,...,13$.

\begin{figure}
\centering
\includegraphics[scale=0.4]{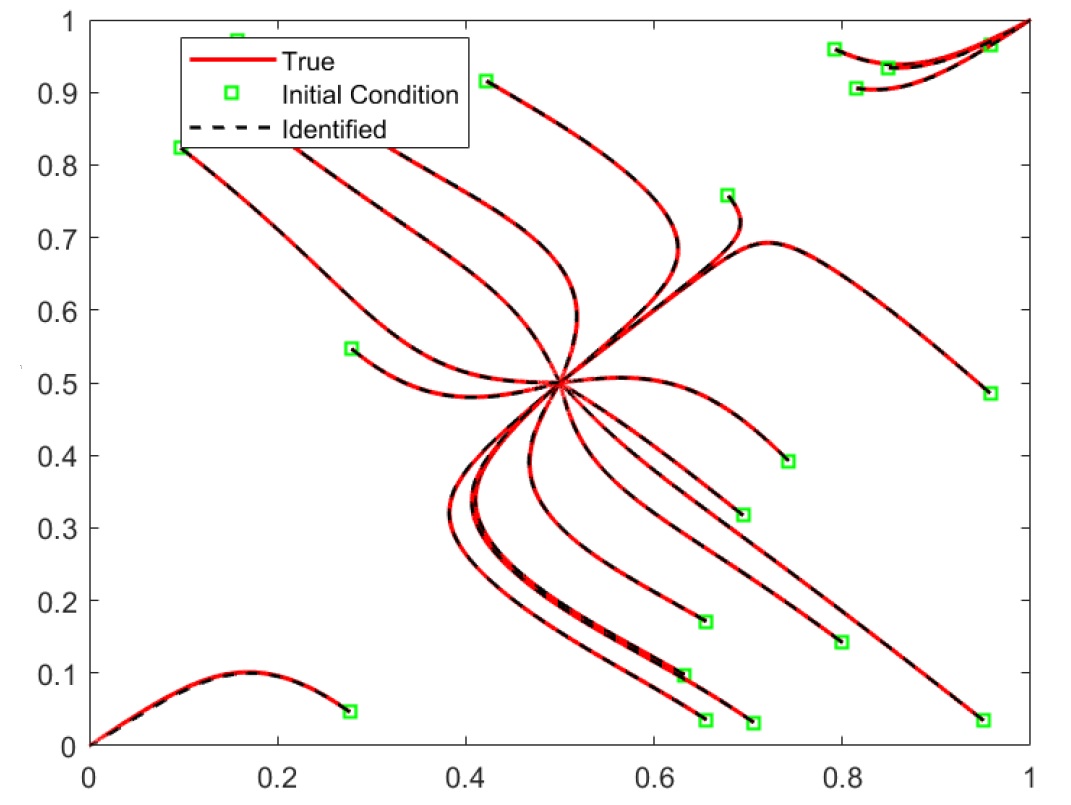}
\caption{Trajectories of the true Equations~\eqref{smoothed-step-system} and trajectories of the  approximation of these equations with the System~\eqref{smoothed-step-approxi-sytem}. There are 3 basins of attraction with the attracting stationary points $(0,0)$, $(.5,.5)$ and $(1,1)$. The trajectories' initial conditions are shown as green squares.}
\label{pic-flow-steps}
\end{figure}

The result of the LASSO regression is the system of the differential equations
\begin{equation}\label{smoothed-step-approxi-sytem}
\left\{
\begin{array}{l}
\begin{aligned}
x' =& -.2669x^5 +.2845y^5 -.3128\sin(3x)+.2479\sin(3y)\\
     & +.3808\cos(3x) -.2753\cos(3y) -.0977\cos(5x)-...\\
y' =& -.2669y^5 +.6149\sin(3x)-.3128\sin(3y)-.7437\cos(3x)\\
     & +.3808\cos(3y)+.9624\cos(5x) -.0977\cos(5y)-...
\end{aligned}
\end{array}
\right. 
\end{equation}
Even though the identified Equations~\eqref{smoothed-step-approxi-sytem} differ from the true Equations~\eqref{smoothed-step-system}, it appears that the approximation of the trajectories is of high precision (Figure~\ref{pic-flow-steps}).

\section{Dynamics of the volume of users on Internet platforms}\label{section-Internet}
In this section, we will use the ideas of the differential equations approximation, discussed in the Sections~\ref{section-vf-vs-flow},~\ref{section-statistics},~\ref{section-approximation-examples}, for the analysis of the traffic on Internet platforms. The estimates of the volume of platform users are important for the platform owners, helping them to price the platforms' services, to predict the moments of high traffic (which occasionally cause platform's crashes) and to understand the long term future of platforms' popularity.

We will focus on the examples of the dynamical system models for the two types of Internet platforms: those that involve interaction similar to the "seller-buyer" relations and those that are similar to the Wikipedia platform. We will analyze tendency and global properties of the volume of users interacting through these platforms. For more detailed analysis and proof of the results stated in this section see \cite{R1} and \cite{R2}.

The platforms discussed in this Section are the {\it two-sided} platforms. They are characterized by two different types of users: renters and home owners (on Homes.mil), sellers and buyers (on Amazon.com), readers and contributors (on Wikipedia.org), males and females (on Match.com), job seekers and employers (on Jobs.com), etc. 

The volume of users in our model is a fraction of all theoretically possible users on each side of the platform. We will denote them $b$  and $g$ (for example, $b$ may represent the fraction of all possible buyers and $g$ may represent the fraction of all possible sellers interacting through the Amazon.com platform) and consider the following system:
$$
\left\{
\begin{array}{l}
b' =V(g) + R(b),\\
g' =W(b) + S(g).
\end{array}\ \ \  b\in [0,1],\ g\in [0,1],
\right.
$$
Here, $R(b)$, $S(g)$ are the same-side network effects, and $V(g)$, $W(b)$ are the cross-side network effects. 

The same side network effect represents the interaction within the same type of users. This effect can be positive ($S:[0,1]\to [0,\infty)$), negative ($S:[0,1]\to (-\infty,0]$, or non-existent ($S\equiv 0$). The same for the $R$ function.

The negative effect is typical for renters and buyers who want to keep costs low, for home owners (on Homes.mil) sellers (on Amazon.com), job seekers (on Jobs.com), males or females (on Match.com) who prefer lower competition. Positive effect occurs between reviewers on Yelp, or gamers who share access tools on gaming platforms. On a platform like Wikipedia, the same-side network effect does not exist (unless there are some "edit wars" between contributors). 

First, we will consider platforms with the negative same-side network effect, such as Homes.mil, Amazon.com, Match.com, Jobs.com. Platforms with the negative same-side network effect we call "seller-buyer" type of platform.
 
Then, we discuss the Wikipedia platform. 

In all examples discussed below, we approximate the same-side network effects with linear functions. 

The cross-side network effect represents user’s preferences and interest in the other type of users. Usually this effect is non-negative (therefore, we assume it here), and  $W(0)=V(0)=0$ (nobody wants to join the platform without presence of the opposite party). Time and range re-scaling allow to assume that $W,V : [0,1] \rightarrow [0,1].$ We will call $W$ and $V$ the attachment functions.

\subsection{Systems with negative same-side network effect}\label{section-seller-buyer-platform}
In this subsection, we discuss the users' traffic model for platforms with the negative same-side network effect, approximated by linear functions $-\epsilon b$ and $-\delta g$:
\begin{equation}\label{neg-same-side-system}
\left\{
\begin{array}{l}
b' = V(g) - \epsilon b,\\
g' =W(b) - \delta g,
\end{array}
\right.
\end{equation}
where 
$$b\in [0,1],\ g\in [0,1],\ \epsilon, \delta \geq 1 \mbox{  and  }V, W: [0,1]\to [0,1],\ \  V(0)=W(0)=0.$$

It is easy to see that each trajectory defined by the Equations~\eqref{neg-same-side-system}, with its initial condition in the unit square $[0,1]^2$, remains in the square and converges to a fixed point (due to negative divergence). In other words, the model is well defined (platform's population does not grow beyond $100\%$) \& has no cyclic trajectories. We will discuss the tendency (convergence to the fixed points) of the trajectories and other qualitative properties of the phase portrait of the system with various attachment functions.

The first example is the power attachment, i.e., 
\begin{equation}\label{eqn-power}
V(g)=g^{\alpha}, \ \ W(b)=b^{\alpha},\ \ \alpha \leq1.
\end{equation}

\begin{figure}
\centering
 \includegraphics[scale=0.3]{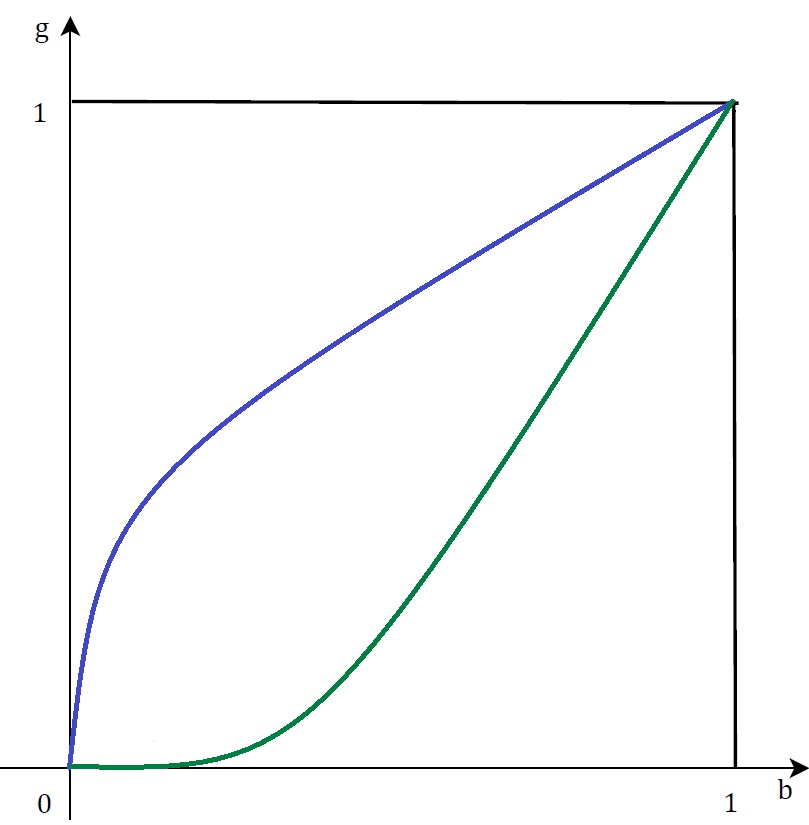}
\caption{The power attachment functions. The fixed points of the dynamical system defined by the Equations~\eqref{neg-same-side-system} with the attachments expressed by the Equations~\eqref{eqn-power} are located at the intersections of the attachment functions: at the origin and at the corner $(1,1)$. This strong attachment attracts all theoretically possible users. Eventually, everybody joins the platform. See Figure~\ref{pic-flow-sqrt} of Section~\ref{section-approximation-examples}.}
\label{fig-power}
\end{figure}

Because fixed points of the dynamical system~\eqref{neg-same-side-system} are located at the intersections of $V$ and $W$, we draw them on the same graph (Figure~\ref{fig-power}). The phase portrait is shown in the Figure~\ref{pic-flow-sqrt} of Section~\ref{section-approximation-examples}. It has one attracting fixed point at $(1,1)$. The quadrant $[0,1]^2\setminus \{0\}$ is the basin of attraction. We can conclude that with the power (very strong) cross-side attachment, all population tends to join the platform eventually.

Many platforms can be characterized by step-function attachments. Here, for simplicity, we show only 3 steps:
\begin{equation*}V(g)=\left\{
\begin{array}{ll}
0, & g\in[0,.25)\\
1/2, & g\in[.25,.75)\\
1, & g\in[.75, 1]
\end{array}
\right.
\end{equation*}
\begin{equation*}W(b)=\left\{
\begin{array}{ll}
0, & b\in[0,.25)\\
1/2, & b\in[.25,.75)\\
1, & b\in[.75, 1]
\end{array}
\right.
\end{equation*}
\begin{figure}
\centering
\includegraphics[scale=0.3]{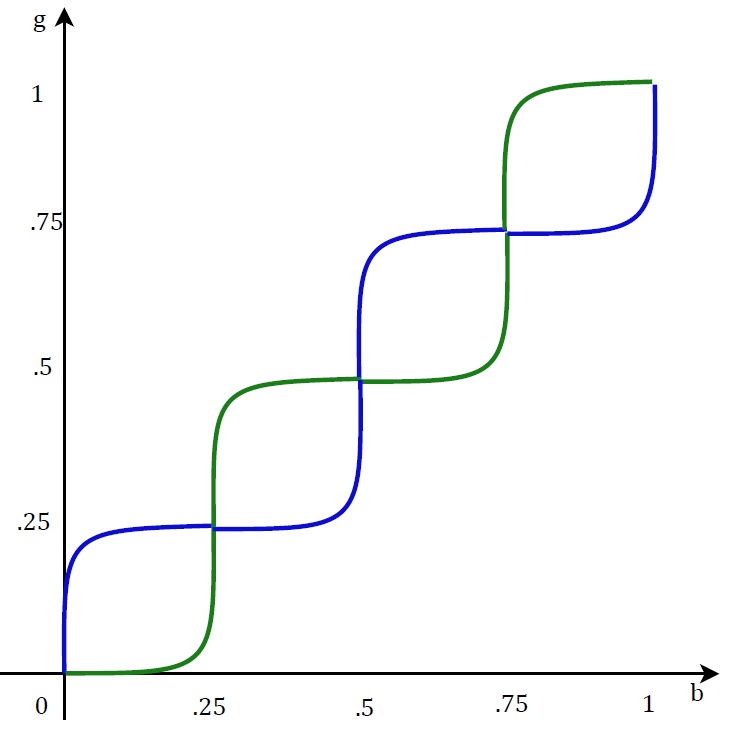}
\caption{Smoothed step attachment (with 3 steps) creates 3 attracting fixed points in the phase portrait of the dynamical system. They are located in the 1st, 3rd and 5th points of intercept of the attachments. The other 2 points of intercept are the hyperbolic fixed points of the dynamical system.}
\label{fixed_pts_smoothed_step_fn}
\end{figure}
These functions are discontinuous and are inconvenient for the analysis. Moreover, it is natural to assume that these steps are smoothly connected (possibly via rapidly growing functions called here $c_1$ and $c_2$). Thus, we will consider smoothed step-function attachments shown in Figure~\ref{fixed_pts_smoothed_step_fn}: 

\begin{equation}\label{eqn-smoothed-step}
V(g)=\left\{
\begin{array}{ll}
0, & g\in[0,.25-\delta]\\
c_1(g), & g\in (.25-\delta, .25 +\delta)\\
1/2, & g\in[\delta + .25,.75-\delta]\\
c_2(g), & g\in (.75-\delta, .75 +\delta)\\
1, & g\in[\delta +.75, 1]
\end{array}
\right.
\end{equation}
A similar equation can be written  for $W(b)$.

\begin{figure}
\centering
 \includegraphics[scale=0.4]{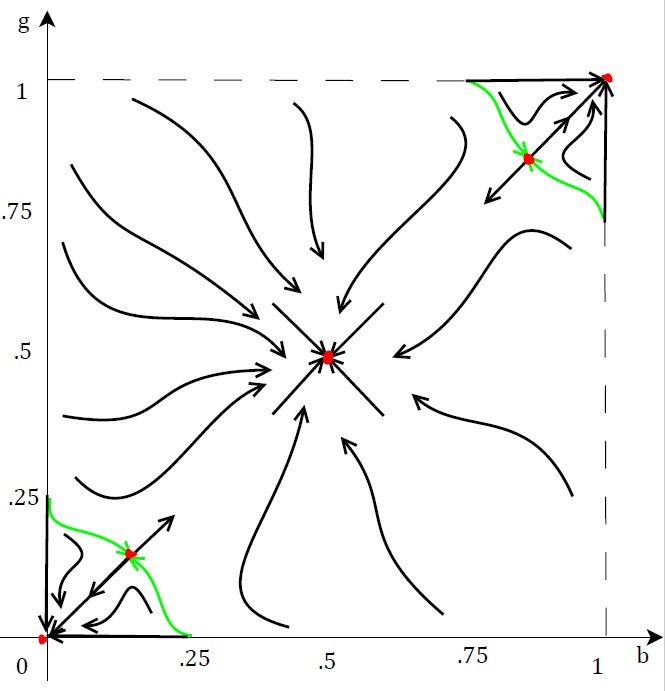} 
\caption{Fixed points are denoted by red dots. Green separatrix, passing through saddle fixed points, separate 3 basins of attraction. Low volume of users (in the low-left basin) eventually extinct. If the volume of users is in the middle basin of attraction, the tendency of the volume is to the middle fixed point. 
If the level of users is in the upper-right basin of attraction, the platform will eventually attract all population. }
\label{smooth_stairs_basins}
\end{figure}

The phase portrait, corresponding to the smoothed step attachment (with 3 steps) is illustrated by the Figure~\ref{pic-flow-steps} of Section~\ref{section-approximation-examples}. It has 3 basins of attraction containing stable fixed points. The basins are separated by separatrix passing through the 2 saddle fixed points. See Figure~\ref{smooth_stairs_basins}. 
\begin{conclusion}\label{conclusion-sell-buy}
The model defined by the Equations~\eqref{neg-same-side-system} with attachments expressed by the Equations~\eqref{eqn-smoothed-step} has the dynamics shown in Figure~\ref{smooth_stairs_basins}. It has the following properties.
\begin{itemize}
\item If the volume of users is low (within the low-left basin), all existing users tend to leave the platform eventually. 
\item If the volume of users is in the middle basin of attraction, the tendency of the volume is the middle fixed point. 
\item If the level of users is high (in the upper-right basin), the platform will eventually attract all population. 
\item The jumps between the basins of attraction happen with the help of external effects. For example, the platform owner can offer some incentives or change the platform's policy to jump from the lower basin to the higher one. 
\end{itemize}
\end{conclusion}

A more general dynamics, defined by the Equations~\eqref{neg-same-side-system},  can be described by the following
\begin{theorem}
Suppose the two-sided platform model described by the Equations~\eqref{neg-same-side-system} is smooth and has finite number of fixed points. Then, they are located at $\{(b,g)\in [0,1]^2: \epsilon b=V(g), \delta g= W(b)\}.$ The fixed points are: stable nodes/saddles/saddle-nodes, and stable spirals (no repelling fixed points -- at least one eigenvalue has negative real part).
\begin{itemize}
\item $(b_0,g_0)$ is a stable node/spiral, if and only if $V'(g_0)\cdot W'(b_0) <\epsilon \delta$;
\item $(b_0,g_0)$ is a saddle, if and only if $V'(g_0)\cdot W'(b_0) >\epsilon \delta$;
\item $(b_0,g_0)$ belongs to center manifold, if and only if $V'(g_0)\cdot W'(b_0) =\epsilon \delta$. Moreover, if $V''(g_0)\cdot W'(b_0) + \frac{1}{\epsilon}( V'(g_0))^2\cdot W''(b_0) \neq 0,$ then one of the branches of center manifold converges to $(b_0, g_0)$, but the other one diverges from this fixed point. 
\end{itemize}
\end{theorem}
The proof of this result can be found in \cite{R2}.

\subsection{Systems without the same-side network effect}\label{section-wiki-platform}
The model for platforms without the same-side network effect can be written as
\begin{equation}\label{eqn-wiki}
\left\{
\begin{array}{l}
b' = V(g),\\
g' =W(b),
\end{array}
\right.
\end{equation}
where 
$$V, W: [0,1]\to [0,1] \mbox{ and } V(0)=W(0)=0.$$
The flow has stationary point at the origin.  Since attachment functions by definition are non-negative, the number of users never declines on such platforms. Also, because $b$ and $g$ never decrease, there are no cycles. Applying these ideas to the general tendency of the popularity of Wikipedia, we can conclude that on the long run, it should increase, if there is no the same-side competition. 

The trajectories of the Equations~\eqref{eqn-wiki} satisfy
\begin{equation}\label{eqn-wiki-solution}
\int V(g)d g-\int W(b) db = 0 .
\end{equation}
 
If $V$  and $W$ vanish not only at the origin, then non-zero stationary points are formed, and the flow of users converges to such point(s) from left and below, while staying within the point's basins of attraction; or the flow escapes the unit square. 

\begin{figure}
\centering
\includegraphics[scale=0.3]{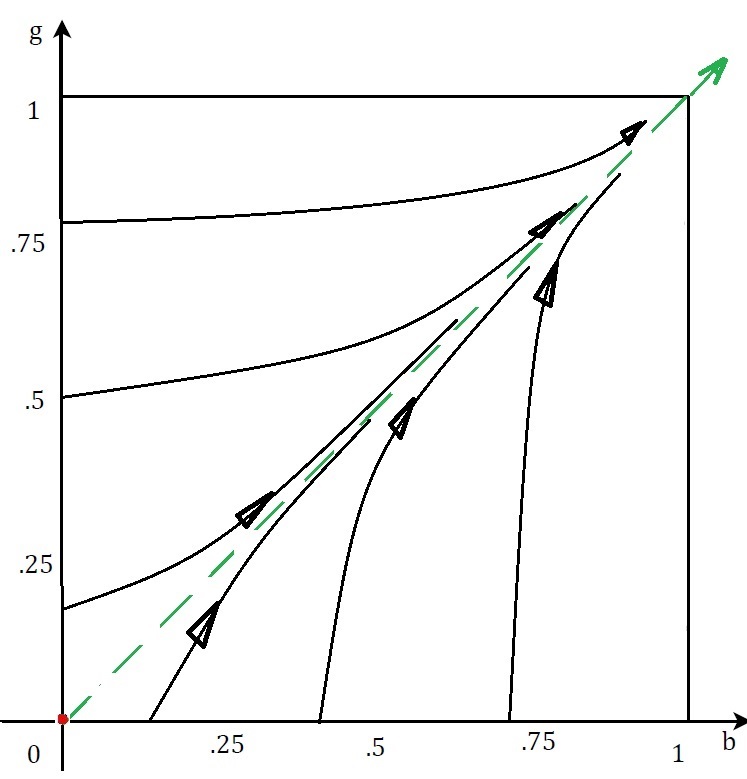} 
\caption{If we assume that negative same-side network effect is not present, the volume of platform users does not decline, unless it jumps to a lower trajectory (due to policy changes, for example). However, if the origin was the only fixed point (as shown here), the flow of users would escape the unit square.}
\label{wiki-flow}
\end{figure}

If $0$ is the only stationary point, or the volume of users is either above the highest stationary point or to the right of the rightmost stationary point, the  trajectories would eventually escape the unit square $[0,1]^2$  as shown in Figure~\ref{wiki-flow}. Thus, we assume that either the negative interaction between the users of the same side must appear at some point and the model changes, or there are some non-zero stationary points (with both, $b$ and $g$ positive coordinates) and the initial volume of user is below and to the left of some stationary point.

Let us come back to the Wikipedia example. The authors of \cite{KBMG},  \cite{HKKR},  \cite{BKM}, \cite{HGMR} discuss issues like "edit wars" between contributors and newcomer retention. They also discuss some new policies, re-interpretations of the community rules and curating according to seniority.
This indicates some competition between contributors and possibly appearance of the  negative same-side network effect. In this case, the system becomes similar to the "seller-buyer" system discussed in the Section~\ref{section-seller-buyer-platform}. Then, incentives provided to the contributors (or some new policy) can move the dynamics into a higher basin of attraction, as discussed in Conclusion~\ref{conclusion-sell-buy}, and increase the volume of contributors and readers.

The recent analysis of \cite{HGMR} shows that the  number of Wikipedia contributors declines. If the "edit wars" are not fundamental characteristics of the Wikipedia platform, they can be viewed as a temporary external effects. In this case, the decline of the volume of users can possibly be explained as jumps from the higher trajectories to lower trajectory, while the fundamental law governing the dynamics has no negative same-side network effects and can be modeled with the Equations~(\ref{eqn-wiki}). Then, the long term tendency of the trajectories is the growth of the number of users.

Let us assume that the Equations~\eqref{eqn-wiki} describe the dynamics of the users' volume of  Wikipedia. As we noticed, in this case $V$ and $W$ must vanish not only at the origin, i.e., at some points $(g_i, b_j)\in (0,1]^2$ each side becomes indifferent to the opposite side:
\begin{equation*}
V(g_i)=0 \mbox{ and }W(b_j)=0.
\end{equation*}

\begin{figure}
\centering
\includegraphics[scale=0.4]{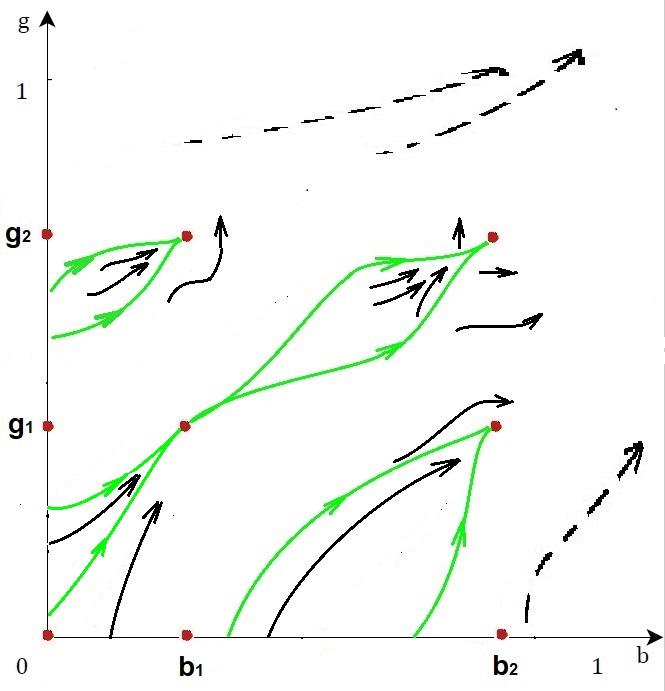}
\caption{If the attachment function $V(g)$ has three zeros (at $0$, $g_1$ and $g_2$) and the attachment function $W(b)$ has 3 zeros (at $0$, $b_1$ and $b_2$), the system has 9 stationary points shown in red color. The flow of users (with some positive volume of users on each side) increases approaching one of the 4 non-zero stationary points staying within some basin of attraction (shown in green), or escapes the unit square. If the flow was above either $b_2$ or $g_2$, (shown with dashed lines) it would escape the unit square.}
\label{fig-wiki-dynamics-with-indifference}
\end{figure}

Then, $(g_i, b_j)$ are non-zero stationary points. Since $V(g)\geq 0 $ on its domain, $g_i$ must be a point of local minimum of $V(g)$; and consequently $V'(g_i)=0$. Similarly, $W'(b_j)=0$. This implies that all non-zero stationary points $(g_i, b_j)$ belong to the  center manifolds, which direct the flow North-East, as shown in Figure~\ref{fig-wiki-dynamics-with-indifference}. Some of the trajectories may be trapped between the center manifolds and converge to the stationary points, while other trajectories escape the unit square. In the latter case, we assume that it takes very long time to escape the unit square (during a reasonably long time the volume of users does not grow beyond 100\%).

\begin{conclusion}\label{conclusion-wiki}
The platform, modeled with the help of Equations~\eqref{eqn-wiki} has the following properties.
\begin{itemize}
\item On the long run, the volume of users does not decrease.
\item A temporary decline of the volume of users may be due to external effects. This corresponds to the jumps from the higher trajectories to the lower trajectories of the governing equations. 
\end{itemize}
\end{conclusion}


\begin{thebibliography}{KH}

\bibitem{BKM} I. Beschastnikh, T. Kriplean, D.W. McDonald {\it Wikipedian Self-Governance in Action: Motivating the Policy Lens}, ICWSM, 2008

\bibitem{BPK} S.L. Brunton, J.L. Proctorb, J.N. Kutzc, {\it Discovering governing equations from data by sparse identification of nonlinear dynamical systems,} {Proceedings of the National Academy of Sciences,} {\bf 113}, (2016), 3932--3937.    

\bibitem{CM} Cowpertwait, P. S. P., \& Metcalfe, A. V. (2009) Introductory time series with R. New York: Springer-Verlag

\bibitem{EPA} T.R. Eisenmann, G. Parker, M. van Alstyne, {\it Strategies for Two-Sided Markets,} Harvard Business Review, {\bf 84(10)}, (2006).

\bibitem{HKKR} A. Halfaker, A. Kittur, R. Kraut, J. Riedl, {it A Jury of Your Peers: Quality, Experience and Ownership in Wikipedia} (2009). WikiSym Article 15, 10 pages. DOI:10.1145/1641309.1641332


\bibitem{HGMR}  A. Halfaker, R.S. Gieger, J. Morgan, J. Riedl {\it The Rise and Decline of an Open Collaboration System: How Wikipedia's reaction to sudden popularity is causing its decline} (2013) American Behavioral Scientist, {\bf 57(5)} 664-688, DOI:10.1177/0002764212469365 

\bibitem{KBMG} T. Krieplean, I. Beschastnikh, D.W. McDonald, S.A. Golder {\it Community, Consensus, Coercion, Control: CS*W or How Policy Mediates Mass Population}, presentation at GROUP 2007.

\bibitem{R1}V. Rayskin,  {\it Dynamics of two-sided markets, Review of Marketing Science}, {\bf 14}, (2016), 1--19.

\bibitem{R2}V. Rayskin,  {\it Users' dynamics on digital platforms}, {\em Mathematics and Computers in Simulation}, {\bf 142}, (2017)

\bibitem{RT}J.-C. Rochet,  J. Tirole, {\it  Platform competition in two-sided markets,} {Journal of the European Economic Association}, {\bf 1(4)}, (2003), 990--1029.

\end{thebibliography}
\end{document}